\documentclass[preprint]{aastex}
\usepackage{epsf}
\shorttitle{Superclustering of Faint Galaxies}
\shortauthors{Tanaka, Yamada, Turner \& Suto}
\def\lesssim{\mathrel{\hbox{\rlap{\hbox{\lower4pt\hbox{$\sim$}}}\hbox{$<$}}}
}

\def\gtrsim{\mathrel{\hbox{\rlap{\hbox{\lower4pt\hbox{$\sim$}}}\hbox{$>$}}}}

\begin{document}

\title{Superclustering of Faint Galaxies in the Field of \\
       a QSO Concentration at $z\sim1.1$ 
\footnote{Based on observations
obtained with the Apache Point Observatory 3.5-meter telescope, which is
owned and operated by the Astrophysical Research Consortium.}
}

\author{Ichi Tanaka and Toru Yamada}
\affil{Astronomical Institute, Tohoku University, Sendai, 980-8578, Japan \\
and \\
National Astronomical Observatory, Mitaka 181-8588, Japan}
\email{itanaka@optik.mtk.nao.ac.jp, yamada@optik.mtk.nao.ac.jp}

\author{Edwin L. Turner}
\affil{Princeton University Observatory, Peyton Hall, Princeton, NJ 08544}
\email{elt@astro.princeton.edu}

\and

\author{Yasushi Suto}
\affil{Department of Physics and Research Center for the Early Universe
(RESCEU)\\
School of Science, University of Tokyo, Tokyo 113-0033, Japan}
\email{suto@phys.s.u-tokyo.ac.jp}

\received{2000 April 13}
\accepted{2000 ??}
\begin{abstract}
We report on a wide-area ($48'\times9'$) imaging survey of faint
galaxies in $R$ and $I$ bands toward the 1338+27 field where an unusual
concentration of five QSOs at $z\sim1.1$, embedded in a larger-scale
clustering of 23 QSOs, is known to exist.  Using a quite
homogeneous galaxy catalog with a detection completeness limit of
$I\sim23.5$, we detect a significant clustering signature of faint red
galaxies with $I>21$ and $R-I>1.2$ over a scale extending to
$\sim20h^{-1}_{50}$~Mpc.  Close examination of the color-magnitude
diagram, the luminosity function, and the angular correlation function
indeed suggests that those galaxies are located at $z\sim 1.1$ and trace
the underlying large-scale structure at that epoch, together with the
group of 5 QSOs. Since the whole extent of the cluster of 23 QSOs
($\sim70h^{-1}_{50}$~Mpc) is roughly similar to the local ``Great
Wall'', the area may contain a high-redshift counterpart of superclusters
in the local universe.
\end{abstract}


\section{Introduction}

Superclusters are the largest recognizable structures in the universe,
and thus retain the imprint of primordial density fluctuations in a
fairly direct manner. Due to several observational difficulties in
detecting such weak clustering over large scales at higher redshifts,
however, only a few superstructures at $z>0.5$ are currently identified
(e.g. Connolly et al. 1996; Lubin et al. 2000).  \citet{Oort81} pointed
out that concentrations of QSOs/AGNs, which are relatively easy to
identity, may be possible sites for such superclusters at higher
redshifts.

 \citet{C88,C89} discovered an unusual concentration of 23
QSOs at $z\sim 1.1$ in $\sim 2.2 \times 2.2 $~deg$^{2}$ field denoted as
1338+27.  Subsequent observations by \citet{H93,H95} and by
\citet{Yamada97} revealed an excess of faint galaxies near
several QSOs in the concentration.
These findings suggest either that the QSOs are located in isolated
rich associations of galaxies or that those QSOs and the galaxies
trace the underlying large-scale structure in a similar manner. While
either interpretation has an interesting implication for the origin
and evolution of spatial biasing of QSOs and galaxies at high
redshifts, previous observations are not able to discriminate between
the above two possibilities due to the limited size of the observed
fields.  We carried out a new deep optical imaging observation of
galaxies in the $48' \times 9'$ region including the highest density
peak of the QSO concentration at $z=1.1$, and found that there are
several concentrations of galaxies whose color and magnitude are
consistent with old galaxies at $z\sim1.1$. While they do not seem to
be directly associated with any particular QSO, they show clustering
over the entire QSO group in our observed area.  This indicates that
both those galaxies and QSOs are the tracers of much larger-scale
structure (possibly even as large as the whole cluster scale of 23
QSOs), perhaps similar to the ``Great Wall'' in the local universe
\citep{dLGH86}.

The rest of this paper is organized as follows.  After briefly
describing our observations and data reduction in \S 2, we present the
results of an object finding analysis and the resulting density map (\S 3.2).  The
color-magnitude diagram and the luminosity function for the detected
high-density regions are considered in \S 3.3. A further analysis by
angular correlation functions for the whole survey region is described
in \S 3.4. Finally we summarize our findings and discuss their
interpretation and implications in \S 4. We assume the Hubble constant
$H_0=50$km/s/Mpc, the density parameter $\Omega_0=1$, and the
cosmological constant $\Lambda=0$ throughout the paper.

\section{Deep imaging of galaxies near the 1338+27 field}

\begin{figure}
\begin{center}
\leavevmode \epsfxsize=15cm \epsfbox{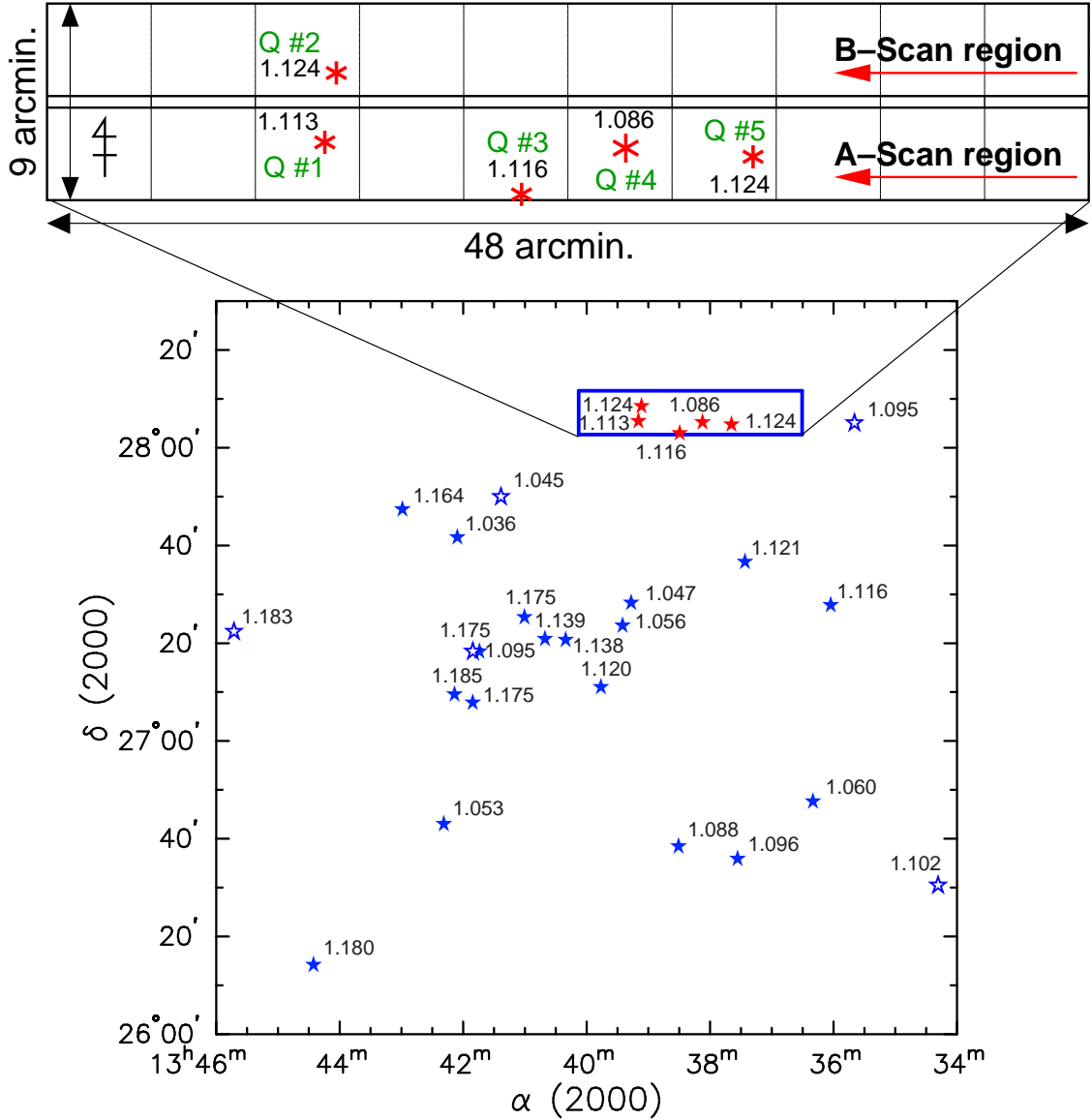}
\end{center}
\figcaption{{\it Top}: The configuration of our $48'\times9'$ TDI survey 
region in the 1338+27 field. The location of the five QSOs with their 
redshifts are shown by asterisks. B2 1335+28, the only radio-loud QSO 
in the area, is ``Q\#4''. The ``A-Scan'' region is characterized by the 
fact that the area contains four of the five QSOs. The whole scan-length 
of 48' corresponds to $24.7 h_{50}^{-1}$ Mpc at $z=1.1$. 
{\it Bottom}: Wide-area view of the concentration of 23 $z\sim1.1$ QSOs 
in the 1338+27 field. The original members of the QSO cluster identified 
by \citet{C89} are shown by filled stars, whereas some other QSOs at 
$1.0<z<1.2$ in the area (their redshifts are determined and supplied to 
the NASA Extragalaxtic Database (NED) after the publication of the paper)
are represented by open stars with their redshifts. 
Our survey field is indicated by the box.
\label{fig:area}
}
\end{figure}

We observed the $48' \times 9'$ region around the highest density peak
of five QSOs in the 1338+27 field (Fig.~2 of Crampton et al. 1989) in
Kron-Cousins--like {\it R} and {\it I} bands on Feb 23 to 25, 1998
using the SPIcam instrument on the 3.5-meter telescope at the Apache
Point Observatory (APO).  We obtained the images of the two adjacent
scan regions (Fig.~\ref{fig:area}) where the group of the five QSOs at
$z\sim1.1$ \citep{C89} exists; four of them are located in the
A-scan (QSO \#1 at $z=1.113$, \#3 at $z=1.116$, \#4 at $z=1.086$ and
\#5 at $z =1.124$), and the other one in the B-scan (QSO \#2 at
$z=1.124$).  The observations were performed in TDI
(time-delay--and--integrate) mode with a scanning speed of 0.2 times
sidereal, i.e., $3''$/second. We operated the SPIcam in the $2 \times
2$ binning mode, and the resulting pixel size and the field of view of
each frame are 0.282$''$ and $4'.78 \times 4'.78$, respectively.  The
net integration time for each object in a scan is thus 96 seconds, and
we repeated the scans of both A and B regions five times in each band. 
The sky was
almost clear and photometric during the exposures and the seeing was
$1.2'' \sim 1.5''$. Details of the reduction procedures are described
in Tanaka (2000).

After flat-fielding orthogonal to the scanning direction, we
obtained $R$ and $I$ images with superior homogeneity across the whole
area. Object detection and photometry used the SExtractor
package ver 2.0 \citep{BA96}. We executed detection on the $R$ and
$I$ combined frame with a detection threshold of $2.1\sigma$ and with
a $3\times3$ top-hat convolution kernel.  We adopted the SExtractor
``MAG\_BEST'' output as the ``total'' magnitude of objects, while we
measured the color of objects on the seeing-matched registered frames
using a fixed circular aperture with $2.8''$ diameter.  We performed the
photometric calibration using our calibrated deep images taken with
the Issac Newton Telescope \citep{Yamada97,Tanakaetal00} whose
observed field significantly overlaps that of the current survey.
Unsaturated and compact objects (mostly stars) detected in both images
were used for the photometric calibration.

\begin{figure}
\begin{center}
\leavevmode \epsfxsize=15cm\epsfbox{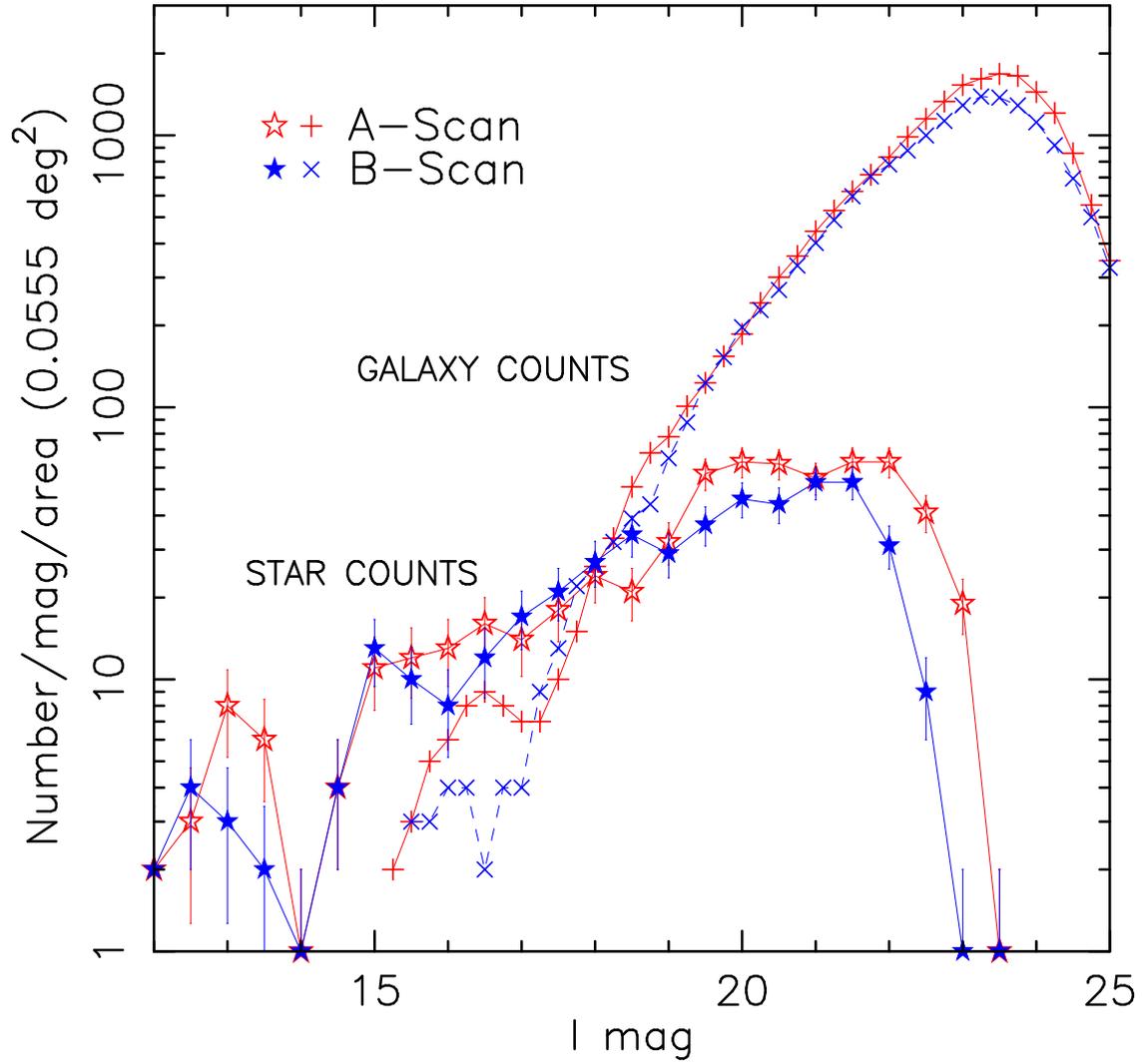}
\end{center}
\figcaption{Results of the star counts (open and filled stars) 
and the galaxy number counts (pluses and crosses) for the A- \& B-Scan 
regions.
\label{fig:nm}
}
\end{figure}

In Figure~\ref{fig:nm}, we show the galaxy number counts as well as
the star counts in each scan region.  We classify objects as ``stars''
when SExtractor the ``stellarity'' index is over 0.90.  Figure~\ref{fig:nm}
shows that the star-galaxy classification is mostly complete up to
$I\sim22$.  In total, we have detected 10370 objects among which 584
objects are classified as stars and removed from the analysis below.

Although the star counts in the A-Scan region are systematically higher
than those in the B-Scan below $I\sim22$ mag., which may be due to the
slightly different seeing size, the effect of the mis-classification
on our analysis below is estimated to be negligible since the
source counts at such faint magnitude are dominated by galaxies.

The ``turn-around'' magnitude in the galaxy number counts is around
$I\sim23.5$ for both the scans. The good agreement of the counts in
the two scan regions indicates that the quality of the images is
globally the same.  There is a slight excess of the number of faint
galaxies in the A-Scan below $I = 22$ mag. It is partially due to the
effect of a rich clusters of galaxies near the eastern edge in the A-Scan
\footnote{The estimated redshift of the cluster is $z\sim 0.5$, with
  the overlap of another poorer cluster at $z\approx0.3-0.4$. Both
  clusters are not reported previously. We find that the latter
  cluster is the counterpart of a radio source 7C 1337+2818 (Waldram
  et al. 1996). The FIRST radio image (at the URL {\tt
    http://sundog.stsci.edu/}) shows that three faint sources are
  closely located at coordinates and the two western sources may
  be associated with the cluster at $z\sim 0.5$.} and partially due to
the population of galaxies studied in this paper.

Since we focus on the difference between the A-Scan (QSO clustered
area) and the B-Scan (comparison field) regions in the analysis below,
we carefully compared the detection efficiency of the two scans using
simulations of artificial objects (Tanaka 2000) and confirmed that the
object detection in both scans is quite homogeneous up to the nominal
completeness limit of $I\sim23.5$.  The photometric accuracy is
evaluated as $\approx 0.2$ mag at $I \sim 22.5$ mag and $\approx 0.5$
mag at $I \sim 23.5$ mag (see errorbars in Fig.~\ref{fig:cmdiag}), and
is almost the same in each scan.  Magnitudes of objects in the overlapped
region of the A-Scan and the B-Scan were also examined. They agree
with an r.m.s.  of $\lesssim0.05$ magnitude.

\section{Detection of clustering of faint galaxies in the field}

\subsection{Expected galaxy colors and magnitudes at $z=1.1$}

\begin{figure}
\begin{center}
\leavevmode \epsfxsize=15cm\epsfbox{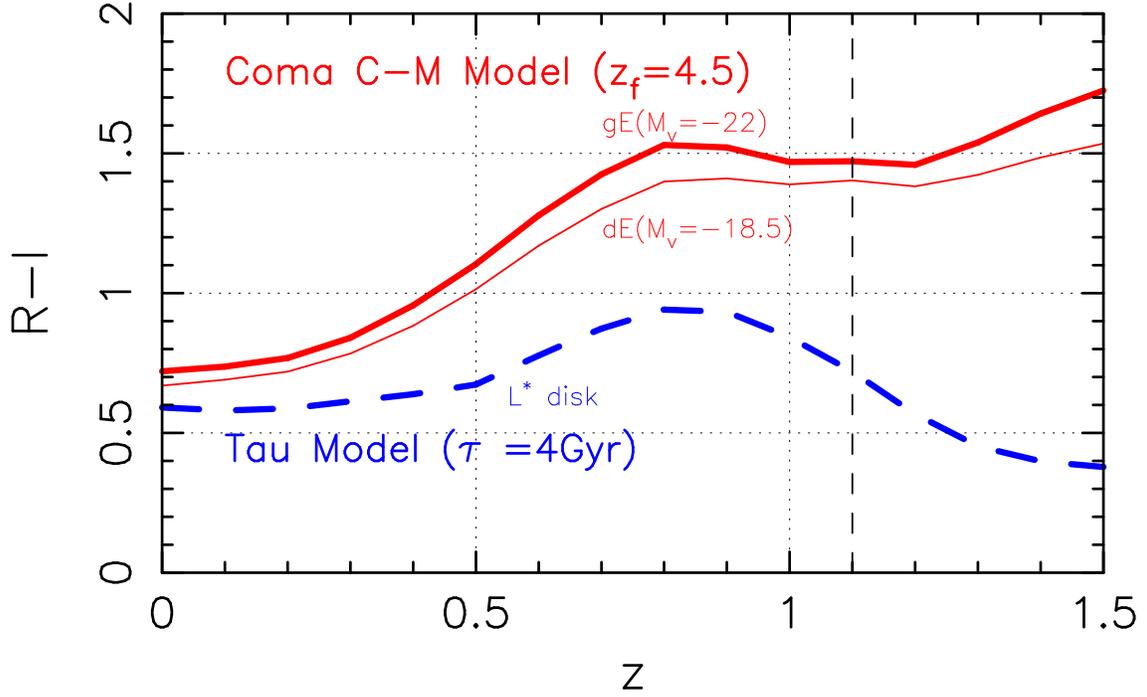}
\end{center}
\figcaption{Predicted evolution of the $R-I$ color for model galaxies.
Thick and thin solid curves correspond to the passively-evolving
giant and dwarf elliptical models with $M_{V}$ of each
$-22$ and $-18.5$ at the age of 12~Gyr (we
assume the formation redshift of $z_{f}=4.5$) calibrated for the
color-magnitude relation of the Coma cluster.
Dashed curve indicates the model galaxy with
star-formation rate $\propto \exp(-t/\tau)$ and $\tau=4$ Gyr with the
formation epoch of $z_f=4.5$.
\label{fig:modelcolor}
}
\end{figure}

Since we do not have the redshift of each identified object, the
projection along the line of sight significantly weakens the signature
of the possible clustering of objects at $z\sim 1.1$. Thus we select
candidates in the corresponding redshift range according to their
color and magnitude.  Figure~\ref{fig:modelcolor} plots the color
tracks of model galaxies as a function of redshift calculated on the
basis of a spectral synthesis model by \cite{KA97}.  In the figure,
solid lines labeled as ``Coma C-M model'' indicate the models of
passively-evolving old galaxies. They are calibrated by the
color-magnitude (C-M) relation of the Coma cluster and are known to
reproduce the evolutionary trend of the cluster early-type galaxies
up to $z \sim 1.2$ \citep{BKT98,SED98,KYT00_cm}.  The dashed line
labeled ``Tau Model'' corresponds to a model with star formation
rate $\propto \exp(-t/\tau)$ and $\tau=4$ Gyr, which may be
appropriate for late-type disk galaxies \citep{Lilly98}.

Figure~\ref{fig:modelcolor} indicates that the color criterion $1.4\lesssim 
R-I \lesssim1.5$ is suitable to photometrically select passively-evolving old
galaxies at $z \gtrsim 0.7$. We also calculate the apparent magnitude
for the ``gE'' model galaxy ($M_{V}=-22$ mag. at 12-Gyrs old or
$z=0$). It monotonically decreases and becomes $I=21.5$ mag at $z=1.1$.
Considering these models as well as the photometric uncertainties, we
finally set two photometric criteria, $1.2<R-I<1.6$ \& $I>21$, to 
extract cluster galaxies at $z\sim1.1$ from the whole sample.  

We note that there are several observational supports for the
criterion that we adopt here.  \citet{Tanakaetal00} show that the
brightest galaxy seen in the color-magnitude sequence of the cluster
around B2~1335+28 at $z\sim1.1$ (Q\#4 in Fig.~\ref{fig:area}) has
$I=21.4$ and $R-I=1.44$. The color and magnitude of the
spectroscopically-confirmed brightest galaxies in the two clusters at
$z\sim1.26$ and $1.27$ studied by \citet{Rosati99} would have $R-I
\approx 1.5$ and $I\sim20.5$ if they were at $z=1.1$.  Star-forming
galaxies at $z\sim1$ studied by Le~F\`ebre et al. (1994) have
magnitudes $I>21$ and converted colors of $R-I\gtrsim1$.

\subsection{Surface density map of the observed area}

The {\it Top} panel of Figure~\ref{fig:map} shows the map of
identified objects satisfying the criteria $1.2<R-I<1.6$ and
$21<I<23.5$. The faintest end, $I=23.5$ is the nominal completeness
limit of the data. While this map already suggests that the clustering
of those galaxies is weakly correlated with the locations of the QSOs
as a whole, this feature is visually more significant from the density
map (the {\it Bottom} panel).  The latter map is constructed from
galaxy number counts on regular meshes whose size is $100\times100$
pixels$^{2}$ $= 28''\times28''$ with additional Gaussian smoothing with a
rms of 100 pixels to ``clean'' the small-scale clustering in the map (see
Tanaka 2000 for further details).

The most distinctive excess in the bottom panel is at around QSO \#4.
In fact this clump corresponds to the cluster at $z \sim 1.1$
identified previously by \citet{Yamada97} and \citet{Tanakaetal00},
and its richness is at least between those of the Virgo and the Coma clusters.
For
reference, the expected number of $R\geqq0$ clusters at $0.7<z<1.3$ in
the area is less than unity at this depth \citep{P96}.

To check whether these red galaxy clumps are consistent with clusters
at $z=1.1$, we identify the most significant ones ($\gtrsim2.5
\sigma$) and then closely examine the color and magnitude distribution
of galaxies in these regions. There are five clumps above this
threshold as labeled in Figure~\ref{fig:map}.

We found that the clump labeled as ``cl\_4" has galaxies
systematically bluer and brighter than the others and therefore concluded
that cl\_4 clump is likely to be a foreground cluster (probably at
$z\sim0.7$) which is not well distinguished from the $z=1.1$ system by
our selection criteria. Indeed, the brightest red galaxy in the clump
has $I = 20.3$. We thus neglect this clump in all further analysis.

Assuming that the remaining four clumps of red galaxies are located at
the same redshift of $z=1.1$ (see next subsections for the further
discussion), we estimated their richness.  Here we used the $N_{0.5}$
parameter of \citet{HL91}, the number of galaxies in the magnitude range
between $m_{1}$ and $m_{1}+3$ within a 0.5 Mpc radius around the
cluster.

The results are summarized in Table~\ref{tbl-1}.  Column 2 shows the
results for the whole sample of galaxies after the field correction
using the counts in the whole B-Scan region where no conspicuous
density excess exists.  Poisson errors are assumed. Column 3 shows the
results for the red-galaxy subsample with $0.8<R-I<2.0$.  The color
selection helps reducing the uncertainly of the field correction since
the majority of the cluster galaxies at $z=1.1$ are expected to lie
within this color range (note however that this choice may exclude
some star-forming very blue galaxies: see Fig.~\ref{fig:modelcolor}).
Column 4 is the estimated Abell richness class based on the
calibration by \citet{HL91}. Note that the magnitude range is very
close to our completeness limit, so the actual $N_{0.5}$ value could
be slightly larger than those derived from our data.

Table 1 shows that those red clumps generally have Abell richness
class $\sim 0-1$ and may be relatively poor clusters except for cl\_2.
Since cl\_2, near QSO\#4, shows fairly lumpy and extended structure, we
also estimate $N_{0.5}$ in the area centered on the QSO (the third
row).  The area does not overlap with cl\_2 region, but the count is
still large.  These results agree with those in \citet{Tanakaetal00}
based on deeper images of this small region.

\subsection{Color-magnitude relation and luminosity function
of the detected galaxy clumps}

In order to make sure that the visual clustering pattern in
Figure~\ref{fig:map} is indeed associated with galaxies located at $z
\sim 1.1$, we have to show that their colors and magnitudes are
consistent with those expected for the galaxies at that redshift.

The combined C-M diagram is plotted in Figure~\ref{fig:cmdiag}. All
the galaxies located on the $500\times500$ pixel$^{2}$ area
($=1.2\times1.2 h_{50}^{-2}$ Mpc$^{2}$) around the four clumps defined
in the previous section are shown in the upper panel. Galaxies in the
richest clump cl\_2 are shown in open squares, while the others are indicated
by
filled circles. A distinctive sequence around $R-I=1.4-1.6$ is clearly
seen in the diagram, which shows excellent agreement with the
expected color of passively-evolving old galaxies at $z\sim 1.1$ (the
horizontal line).  Even if we exclude the contribution of the richest
clump (the lower panel) the red sequence can be recognized clearly. We
also see that there is no systematic difference in the colors of the
red sequence among these three clumps.

Next we examine the luminosity function (LF) of the galaxies in the
clumps. The LFs of $I$-band selected ``red'' galaxies in the CFRS
\citep{Lilly95} are fitted by the Schechter form (Schechter 1976) with
$M^{*}_{B}=-22.89$ for a $0.75 < z < 1.00$ sample.  The extrapolation of
the Virgo cluster LF ($M^{*}_{B}=-21.4$: Sandage et al.1985; Colles
1989; Lumsden et al.1997) assuming passive evolution predicts
$M^{*}_{B} = -22.9$ at $z=1.1$. Note that De Propris et al.(1999) and
Kajisawa et al.(2000a) showed that the $m^{*}$ value in near-infrared
$K$-band LFs for clusters at $z<1$ is fully consistent with passive
evolution.

We now wish to compare these $M^{*}_{B}$ values with 
those of the galaxies in the four red
clumps detected in our survey. Assuming that they are all located at the
same redshift, we combine the data of all the clumps and fit them to
the Schechter function.  In order to reduce the uncertainty in the
field correction, we again use the ``red galaxy'' sample ($R-I>0.8$)
for the analysis (Note that our applied color criteria is similar to
the criteria used for ``red'' galaxies in the CFRS). The field
correction is made using the counts of the B-Scan region. Since our
data are not sufficiently deep to allow a fit of the faint-end slope
$\alpha$ and $M^{*}$ simultaneously, we fix the slope at $\alpha=-0.9$
following De Propris et al.(1999) and Kajisawa et al.(2000a) and then
calculate the best-fit value of $M^{*}$.

Figure~\ref{fig:lf} shows the results. For comparison, the luminosity
function of galaxies excluding those in the richest clump cl\_2 is
also plotted in open triangles. The counts for each clump are also
summarized in Table~\ref{tbl-2}.  The standard $\chi^{2}$-fit to the
Schechter function yields $m^{*}(I) = 22.18^{+0.7}_{-0.3}$ for the
combined sample.  Using the SED template of the passive evolution
model with 3.25 Gyr old galaxies at $z=1.1$, this translates to an absolute
$B$-band magnitude of $M^{*}_{B}=-22.99$.  Our $M^{*}_{B}$ value
agrees with the result of CFRS as well as that expected from the local
cluster LF, which clearly supports the idea that these clumps of
red galaxies are located at $z \sim 1.1$, the same redshift as the QSO
group.

\subsection{Angular correlation function}

Lastly we investigate the clustering of the faint galaxy population more
quantitatively using the angular two-point correlation function
$\omega(\theta)$.  For this purpose, we construct several
color-selected subsamples, and compute the auto-correlation functions
of each subsample, $\omega_{ii}$, and the cross-correlation functions
between two different subsamples of galaxies, $\omega_{ij}$.  We use
the Landy \& Szalay (1993) estimator for the former,
$\omega_{ii}(\theta)=(DD_{ii}-2DR_{ii}+RR_{ii})/RR_{ii}$, and
$\omega_{ij}(\theta)=DD_{ij}/RR_{ij} -1$ for the latter, where $DD$,
$RR$, and $DR$ are the data-data, random-random, and data-random pairs
of separation between $\theta - d\theta/2$ and $\theta + d\theta/2$,
and the subscripts $i$ and $j$ refer to different subsamples.
Error-bars are calculated by the standard boot-strap resampling method
with 100--500 random resamplings (e.g. Barrow, Bhavsar, \& Sonoda
1984). We did not apply the integral constraint correction or the
correction for star-galaxy misclassification (e.g. Postman et al.
1998), since our analysis is based on the inter-comparison of signals
between the A-Scan and the B-Scan regions. We note that both corrections tend
to enhance the amplitude of the measured correlation signal, albeit
only slightly.

The resulting angular correlation functions are plotted in
Figure~\ref{fig:correlation}.  Solid and dashed lines correspond to
the results for the A-Scan and the B-Scan, respectively.
While the clustering amplitudes for such faint galaxies are generally
very weak, we see a strong signal for {\it red} ($1.3<R-I<2.0$)
galaxies in the A-Scan region (the middle-left panel in
Figure~\ref{fig:correlation}).  This is in marked contrast with {\it
  blue} galaxies ($0.9<R-I<1.3$) which do not show appreciable
auto-correlation anywhere.  A similar color-selected correlation study
by \citet{Woods97}, whose sample has a similar depth and width to our
data, showed that $\omega(\theta)$ for the {\it red} ($R-I>1$ or
$>1.5$) galaxy subsample is $\sim0.005$ at $\theta=35$ arcsec.  Our
result for the {\it red} ($1.3<R-I<2.0$) galaxy subsample in the
B-Scan region and the {\it blue} ($0.9<R-I<1.3$) galaxy subsample
agrees with their result within the errors. In contrast, our detected
signal for the A-Scan region is an order of magnitude stronger than that
in such typical fields.

Since the clustering signal for red galaxies becomes significantly
weaker when we exclude the contribution of the four clumps (by removing
the galaxies within a 1.2-Mpc radius for the richest cl\_2 and a 0.6-Mpc
radius for the others), it is indeed caused by the existence of these
high-density clumps. The thin dot-dashed lines in the middle right panel
show the systematic trend of the decreasing clustering signal as
the clumps are removed one-by-one.  A large decrease occurs when we
exclude the cl\_2 area,  but a  still significant clustering amplitude
remains from the contribution of the other three clumps.

We also detect a slightly weaker but robust cross-correlation between
blue and red galaxies, which is insensitive to the presence of the four
clumps. Although we examined cross-correlations for the various
color ranges, we did not find a similar signal between the other
subsamples. Since the cross-correlation persists between the $0.9 < R-I
< 1.1$ and $1.4 < R-I < 2.0$ subsamples, this cannot be ascribed to
scatter of the observed color (due to photometric errors) of the red
objects. This indicates that the A-Scan region contains a clustering of
very {\it red} galaxies and somewhat {\it bluer} galaxies. The color range 
showing the signal corresponds to early-type ellipticals ($1.3<R-I<2.0$)
and the mildly star-forming normal spiral galaxies ($0.9<R-I<1.3$)
at $z=1.1$. One would normally expect that the majority of galaxies 
with bluer colors would
be in the foreground of these red galaxies (see
Fig.~\ref{fig:modelcolor}) and would not yield a significant
cross-correlation signal. Further analysis indicates
that this surprising signal may not be localized in any one area
of the A-Scan region but rather is due to a diffuse structure extended over the
whole region (Tanaka 2000).

\section{Conclusions and discussion}

\begin{figure}
\begin{center}
\leavevmode \epsfxsize=15cm\epsfbox{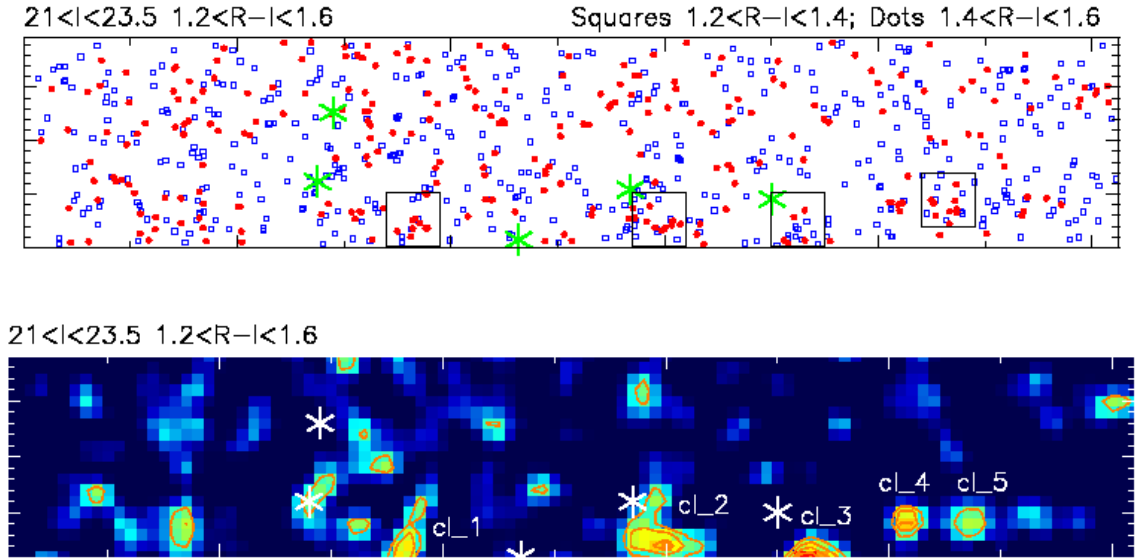}
\end{center}
\figcaption{{\it Top:} Positions of the identified
 galaxy candidates ($21<I<23.5$) at $z\sim1.1$ in the whole
 survey area ($48'\times9'$). The position of each QSO is shown by
 asterisks. Dots and Boxes indicate
galaxies with 1.4$<${\it R$-$I}$<$1.6, and 1.2$<${\it
 R$-$I}$<$1.4, respectively. Inset boxes indicate areas that are
 used for the analysis in Fig.~5. {\it Bottom:} Smoothed density map of the
 red galaxies with $21<I<23.5$ and $1.2<R-I<1.6$.  The figure illustrates
 the amplitude of the density contrast in units of its rms value
 $\sigma$, and the plotted contours indicate the $2\sigma$,
 $3\sigma$, $4\sigma$ levels.
\label{fig:map}
}
\end{figure}

\begin{figure}
\begin{center}
\leavevmode \epsfxsize=15cm\epsfbox{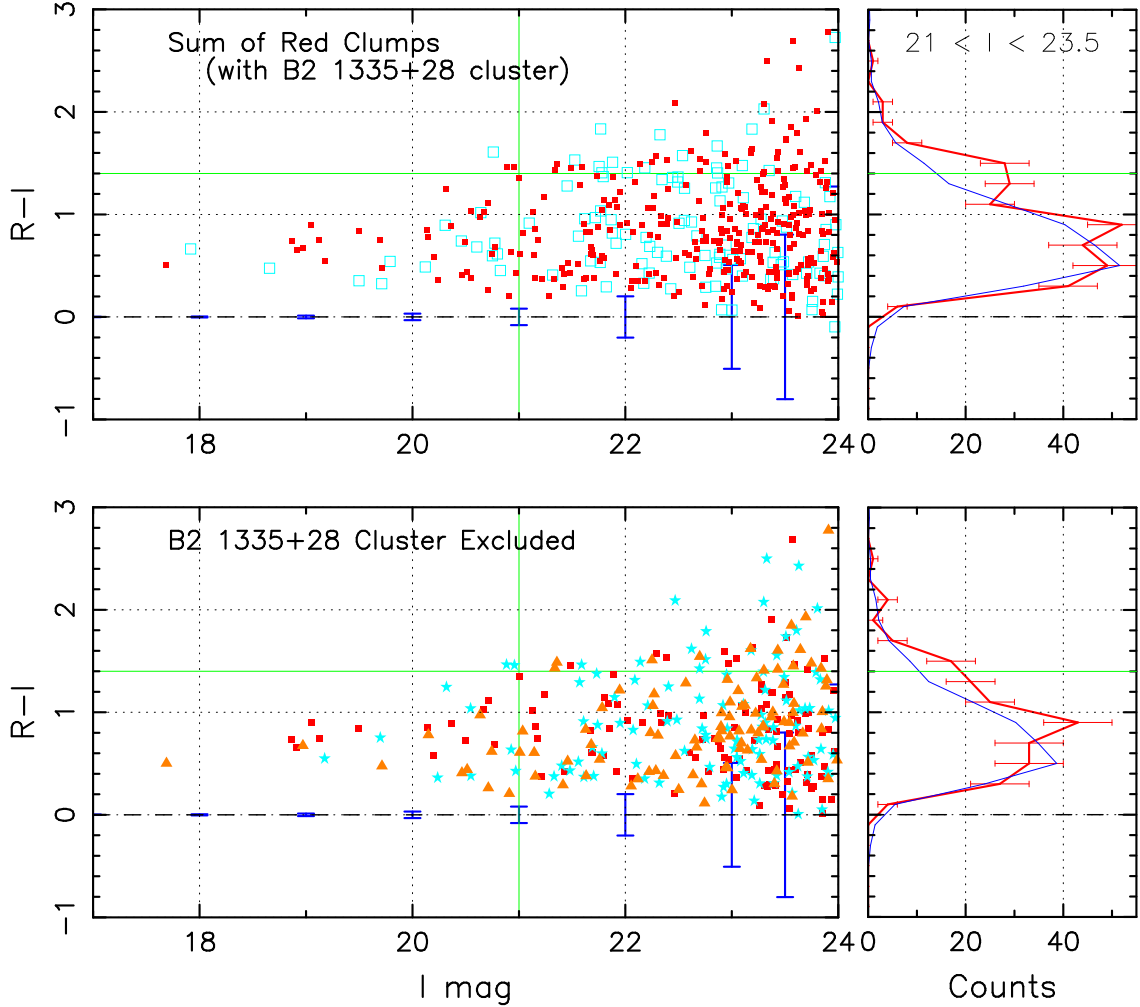}
\end{center}
\figcaption{{\it Top:} Color-magnitude diagram for the galaxies in the four
$2\arcmin.4 \times 2\arcmin.4$ regions around the
clumps exceeding $2.5\sigma$ density in Fig.~\ref{fig:map} (cl\_1, cl\_2,
cl\_3, \& cl\_5). Open squares indicate galaxies in cl\_2, a known rich
cluster near B2 1335+28 (Q \#4). The position of $R-I=1.4$ and $I=21$ 
are marked by thin lines for clarity. The error-bars plotted on the $R-I=0$ 
line are estimated for galaxies with $R-I=1$ from our simulation.
The right panel shows the projected 
color distributions for galaxies with $21<I<23.5$ with $\Delta(R-I)=0.2$ 
bins (thick lines with errorbars). The histogram with thin lines indicates 
the expected average counts obtained using the whole B-Scan region. 
Poisson error-bars are assumed in the histogram.
{\it Bottom:} The same as the upper panel, but the cluster 
around B2 1335+28 is excluded and the galaxies in each clump are shown 
in different symbols (squares for cl\_1, stars for cl\_3, and triangles 
for cl\_5). 
\label{fig:cmdiag}}
\end{figure}

\begin{figure}
\begin{center}
\leavevmode \epsfxsize=15cm\epsfbox{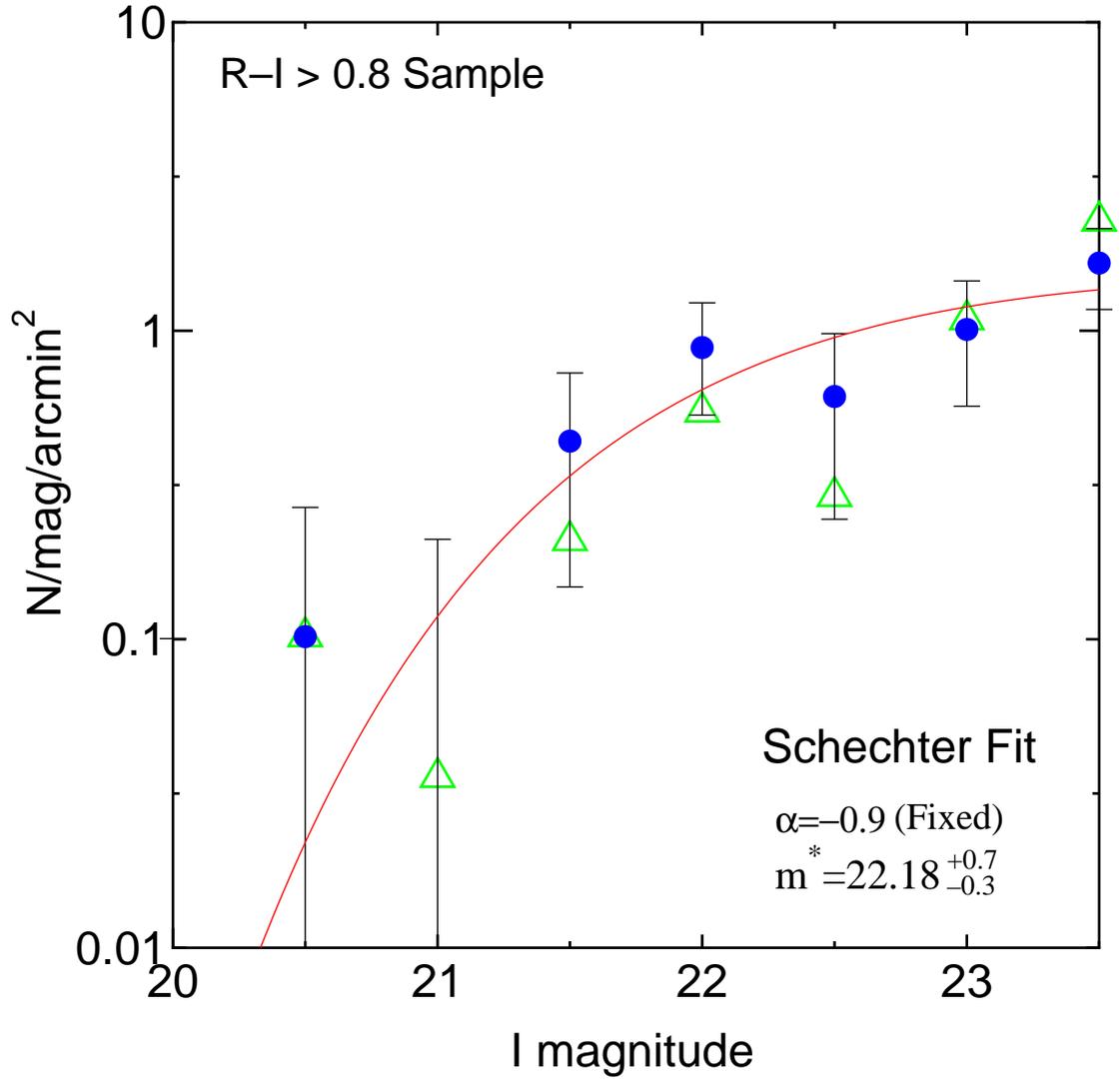}
\end{center}
\figcaption{The luminosity function of red ($R-I>0.8$) galaxies in the
  fields around four $>2.5\sigma$ over-density regions. Filled circles
  indicate results with a rich cluster around Q \#4 (B2 1335+28),
  while open triangles correspond to those without it. The solid curve
  indicates the best-fit Schechter function with a fixed slope of each
  $\alpha=-0.9$.
\label{fig:lf} }
\end{figure}

\begin{figure}
\begin{center}
\leavevmode \epsfxsize=15cm\epsfbox{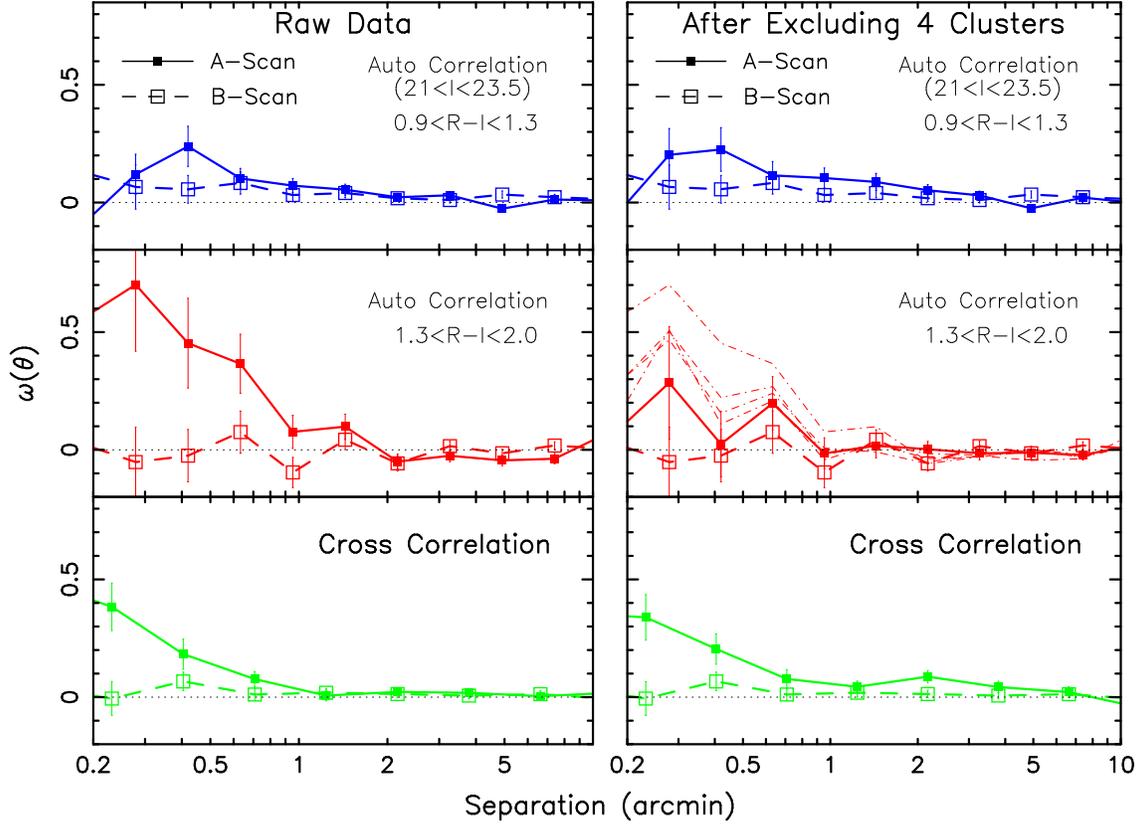}
\end{center}
\figcaption{Angular correlation function of faint galaxies in our
survey. Solid and dashed lines correspond to results for A-scan and
B-scan, respectively. Left and right panels show the results with and
without the four galaxy clump regions, respectively. The top and middle
panels plot the auto-correlation of {\it blue} ($0.9<R-I<1.3$) and {\it
red} ($1.3<R-I<2.0$) galaxies, respectively, while the bottom panels
plot the cross-correlation of {\it blue} and {\it red} galaxies.
Thin dot-dashed lines in the middle-right panel indicates the decreasing
trend of auto-correlation signal with excluding four $>2.5\sigma
$ clumps one by one.
\label{fig:correlation} }
\end{figure}

We have detected a strong clustering signal for the faint red galaxies in the
$48'\times9'$ region around a tight group of five QSOs in the 1338+27
field.  This structure does not seem to be confined to regions
adjacent to each QSO (see Fig.~\ref{fig:map}), but rather
preferentially exists only in the A-Scan region where four of the five
QSOs lie; it might even extend over the entire 23 QSO concentration at
$z\sim1.1$.  The color-magnitude relation, the luminosity function and
the angular correlation functions of those faint galaxies consistently
indicate that the clustered galaxies are located at a similar redshift
to the QSO concentration. This suggests that those galaxies together
with the five QSOs may both trace underlying large-scale
cosmic structure at $z\sim1.1$.

The elongated morphology and size of $\sim 2$ by $20h^{-1}_{50}$ Mpc
of the structure is indeed typical of the known superclusters at low
and intermediate redshifts. \citet{Jaaniste98} studied a sample of
42 low-redshift ($z<0.12$) superclusters and concluded that elongated
filamentary morphology with a size of few Mpc by several tens of Mpc
is typical for the poor ($N_{cl}<8$) superclusters. There are
also examples of superclusers at intermediate redshift: The cl0016
supercluster at $z\sim0.54$ identified in Koo et al.(1985) has a thin
sheet-like structure with a size of 62 by 24 by $8h_{50}^{-1}$ Mpc
(Connolly et al. 1996).  The supercluster at $z\sim0.9$ studied by
Lubin et al.(2000) also has an extent of $\sim10h_{50}^{-1}$ Mpc. However, the
$z=1.26$ system investigated by Rosati et al.(1999) has a 
projected scale of only
$\sim2h_{50}^{-1}$ Mpc.

In addition to the strong auto-correlation signal of the red galaxies
in the four highest density clumps, we have also found a weak but
persistent cross-correlation between {\it blue} and {\it red}
galaxies.  It is not caused by the galaxies in the regions of the four
clumps alone but is likely to come from the whole A-Scan region.  We
speculate that if the structure in our surveyed region is really a
supercluster at $z = 1.1$, the significant
cross-correlation between {\it blue} and {\it red} galaxies may
originate from the association of early-type galaxies and relatively
blue disk galaxies in groups distributed between the four richest
clumps of red galaxies. In nearby cosmic structures such an
association is typically seen\citep{PG84,Z98}. Indeed, Davis \& Geller
(1976) also detected a significant angular cross-correlation between
ellipticals and spiral galaxies using a large-area catalog that
contains the Local Supercluster.

Finally, we also briefly discuss the immediate environment of the 
five individual QSOs.
QSO\#4 (B2 1335+28) is the only radio-loud quasar (RLQ) in our field.
Although RLQs often reside in the central regions of clusters (e.g.
Ellingson \& Yee 1994), \citet{SG99} found that several RLQs at $z>1$
are also located in the outskirts of cluster regions. In our case, Tanaka
et al.(2000) showed that QSO\#4 also resides on the northern edge of the
distribution of the cluster galaxies.

A possible association of a low-significance (2 $\sigma$) clump of red
galaxies with a radio-quiet QSO (RQQ) Q\#1 is seen in the density map
of Figure~\ref{fig:map}. Indeed, \citet{H95} argued that there is a
3-$\sigma$ excess of faint (but $I\lesssim22$) blue ($R-I<1.0$)
galaxies in this location. They also detected some emission-line 
galaxy candidates
around QSOs \#1 and \#2. We examined the C-M diagram of the galaxies
around the density peak near Q\#1.  Sequences of galaxies can be
recognized around $R-I=1.1-1.3$ and $R-I\sim 0.7$ but they are too
blue to be old passively-evolving galaxies at $z=1.1$. Furthermore,
the brightest galaxy in the redder C-M sequence, which lies at the
center of the clustering, has $I=19.5$ and therefore is too bright for
a quiescent brightest cluster galaxy at $z=1.1$. Thus, the weak density excess
in the red galaxy distribution near QSO\#1 may be a chance projection of a
foreground cluster, while our data do not rule out the existence of a
group of blue galaxies associated with QSO \#1. There is no
significant density peak coincident with QSO\#2, QSO\#3, or QSO\#5

Thus, all four RQQs seem not to be associated directly with
rich clusters/groups. It is generally believed that RQQs avoid cluster
environments and are fairly isolated \citep{Yee84,boyle93,Teplitz99}.
Our result agrees with the general trend, even though our RQQs are
strongly clustered. When we consider the distribution of the QSOs and
the clumps of red galaxies globally, however, we note that RQQs may
preferentially reside farther into the outskirts of galaxy clusters. Yee
(1990) also reported such a trend based on a study of low-redshift
RQQ environments.

To date there have been only a few other observational attempts to
directly relate the clustering of AGNs and superclusters (Ford
et al. 1983; Longo 1991; Ellingson \& Yee 1994). On the other hand,
several extensive studies searching for large-scale clustering of
QSOs/AGNs have already been performed (e.g. Oort et al. 1981; West
1991; Graham, Clowes, \& Campusano 1995; Komberg, Kravtsov, \& Lukash
1996).  More homogeneous samples of such AGN groups/clusters are likely to be
constructed from the 2dF QSO survey and the Sloan Digitized Sky
Survey. Clearly, more extensive studies that determine the relationship
of such AGN groups and superclusters to underlying large scale structure are of
great importance.

The detected signature of the putative supercluster at $z=1.1$ may be the first
direct indication of the association of QSOs and a high-$z$
supercluster.  Certain confirmation of this picture, however, must
await spectroscopic follow-up observations of the field with
8--10-m-class telescopes. We are currently planning to carry out
multi-slit spectroscopic follow-up observations of galaxies in the A-Scan
region using the Subaru 8.2-m Telescope. We will also extend the
survey area to the whole concentration of 23 QSOs at $z\sim1.1$ using
deep multicolor imaging with the Subaru Prime-focus Camera. These
data should unambiguously reveal the actual distribution of clusters
and galaxies in this postulated $z\sim1.1$ counterpart of the local
``Great Wall'' structure.

\acknowledgments

We thank M. Hattori for useful discussions, A. Arag\'on-Salamanca, K.
Ohta, N. Arimoto, T. Miyaji, and T. Kodama for their kind permission
to use the INT data for our photometric calibration.  We also thank
T. Kodama for allowing us to use his model galaxy SEDs.  We are very
grateful to the staff at APO for kind assistance during the observing
runs.  This research was supported in part by Grants-in-Aid from the
Ministry of Education, Science, Sports and Culture of Japan (07CE2002,
08740181, 09740168) to RESCEU and to T.Y., and by NSF grant
AST98-02802 to E.L.T.  I.T. gratefully acknowledges travel support
to APO from the Hayakawa Foundation.

\clearpage
\begin{deluxetable}{lrrr}
\footnotesize
\tablecaption{Richness of Each Clump. \label{tbl-1}}
\tablewidth{0pt}
\tablehead{
\colhead{ID Name}
&\colhead{N$_{0.5}$\tablenotemark{a}}
&\colhead{N$_{0.5}$(Red)\tablenotemark{b}}
&\colhead{R$_{\bf Abell}$\tablenotemark{c}}
}
\startdata
cl\_1   &  $5 \pm 6$   &  $8  \pm 3$ & $<0$ \\
cl\_2   &  $10 \pm 7$  &  $10 \pm 5$ & 0  \\
cl\_2 (QSO)\tablenotemark{d} & $25 \pm 8$ & $28 \pm 6$ & 2  \\
cl\_3   & $16 \pm 7$ & $14 \pm 5$ & 1  \\
cl\_5   & $8 \pm 6$ & $11 \pm 5$ & 0  \\
\enddata
\tablenotetext{a}{Field-corrected number of galaxies within 0.5 Mpc radius
with $m_{1}<m_{I}<m_{1}+3$.}
\tablenotetext{b}{Estimated using red ($0.8<R\!-\!I<2.0$) galaxy subsample.}
\tablenotetext{c}{Abell richness class based on the calibration by
Hill \& Lilly (1991)}
\tablenotetext{d}{Estimated for the QSO-centered area. Note that the count
may be affected by a foreground galaxy group in the area. It does not overlap
with the cl\_2 area.}
\end{deluxetable}

\begin{deluxetable}{rrrrrccc}
\footnotesize
\tablecaption{Luminosity Distribution of Red ($R-I>0.8$)
Galaxies\label{tbl-2}}
\tablewidth{0pt}
\tablehead{
\colhead{$I_{\bf med}$\tablenotemark{a}}
&\colhead{$N_{\bf 3 cl}$\tablenotemark{b}}
&\colhead{$N_{\bf 4 cl}$\tablenotemark{b}}
&\colhead{$dN_{\bf 3 cl}$\tablenotemark{c}}
&\colhead{$dN_{\bf 4 cl}$\tablenotemark{c}}
&\colhead{$d\rho_{\bf 3 cl}$\tablenotemark{d}}
&\colhead{$d\rho_{\bf 4 cl}$\tablenotemark{d}}
&\colhead{$\rho_{\bf field}$\tablenotemark{e}}
}
\startdata
20 & 3 & 4 & $-1.8$ & $-2.4$ & $\cdots$ & $\ldots$  &0.2219\\
20.5 & 9 & 12 & 1.7   & 2.2  & 0.1020 & 0.1020 &0.3367\\
21 & 14    & 18 & 0.6   & 0.1  & 0.0357 & 0.0051 &0.6161\\
21.5 & 24 & 37 & 3.4   & 9.6  & 0.2092 & 0.4388 &0.9452\\
22 & 35 & 54 & 8.9   & 19.2 & 0.5459 & 0.8827 &1.1977\\
22.5 & 39 & 59 & 4.8   & 13.3 & 0.2908 & 0.6122 &1.5727\\
23 & 65 & 85 & 17.8  & 22.0  & 1.0867 & 1.0102 &2.1696\\
23.5 & 88 & 104 & 37.1  & 36.1  & 2.2704 & 1.6582 &2.3380\\
\enddata
\tablenotetext{a}{The median $I$ magnitude of each bin}
\tablenotetext{b}{The sum of raw galaxy counts within the 0.5Mpc-region
around
each red galaxy clump. $N_{3cl}$ is the result without a rich cluster cl\_2,
and $N_{4cl}$ is that with cl\_2.}
\tablenotetext{c}{The result of counts after field correction. The whole
B-Scan area is used as field data.}
\tablenotetext{d}{Net excess density per arcminute$^{2}$.}
\tablenotetext{e}{``Field'' galaxy surface density from the counts
in the whole B-Scan area.}
\end{deluxetable}

\clearpage

\end{document}